\documentclass[apjl]{emulateapj}
\usepackage{comment}
\usepackage{ifthen}
\usepackage{amsmath}
\usepackage{amsfonts}
\usepackage{amssymb}
\usepackage{multirow}
\usepackage{wasysym}
\usepackage{subfigure}
\usepackage{epsfig}

\newcommand{\obs}{\mathrm{obs}}
\newcommand{\tru}{\mathrm{true}}

\newboolean{emulateapj}
\setboolean{emulateapj}{true}

\newboolean{color}
\setboolean{color}{true}

\shortauthors{Kipping et al.}
\shorttitle{Flicker as a tool for characterizing planets}
\ifthenelse{\boolean{emulateapj}}{
    
}{
    
}

\begin{document}

%% Titlepage
\title {Flicker as a tool for characterizing planets through Asterodensity Profiling
}

%% Authors
\author{
	{\bf D.~M.~Kipping\altaffilmark{1,2},
             F.~A.~Bastien\altaffilmark{3},
             K.~G.~Stassun\altaffilmark{3,4},\\
             W.~J.~Chaplin\altaffilmark{5,6},
             D.~Huber\altaffilmark{7,8},
             L.~A.~Buchhave\altaffilmark{9}
	}
}
\altaffiltext{1}{Harvard-Smithsonian Center for Astrophysics,
		Cambridge, MA 02138; email: dkipping@cfa.harvard.edu}

\altaffiltext{2}{NASA Carl Sagan Fellow}

\altaffiltext{3}{Dept. of Physics \& Astronomy, Vanderbilt University,
                 1807 Station B, Nashville, TN 37235}

\altaffiltext{4}{Physics Dept., Fisk University, 1000 17th Ave. N,
                 Nashville, TN 37208}

\altaffiltext{5}{School of Physics \& Astronomy, University of Birmingham,
                 Birmingham B15 2TT, UK}

\altaffiltext{6}{Stellar Astrophysics Centre, Aarhus University, Ny Munkegade 
                 120, DK-8000 Aarhus C, Denmark}

\altaffiltext{7}{NASA Ames Research Center, Moffett Field, CA 94035}

\altaffiltext{8}{SETI Inst., 189 Bernardo Av., Mountain View, CA 94043}

\altaffiltext{9}{Centre for Star and Planet Formation, University of Copenhagen, 
                 DK-1350, Copenhagen, Denmark}

%% EOF authors

% #####################################################################
%% abstract
\begin{abstract}

Variability in the time series brightness of a star on a timescale of 8\,hours,
known as ``flicker'', has been previously demonstrated to serve as a proxy
for the surface gravity of a star by \citet{bastien:2013}. Although surface
gravity is crucial for stellar classification, it is the mean stellar density
which is most useful when studying transiting exoplanets, due to its direct
impact on the transit light curve shape. Indeed, an accurate and independent 
measure of the stellar density can be leveraged to infer subtle properties of a 
transiting system, such as the companion's orbital eccentricity via 
asterodensity profiling. We here calibrate flicker to the mean stellar 
density of 439 \emph{Kepler} targets with asteroseismology, allowing us to 
derive a new empirical relation given by 
$\log_{10}(\rho_{\star}\,[\mathrm{kg}\,\mathrm{m}^{-3}]) = 5.413 - 
1.850 \log_{10}(F_8\,[\mathrm{ppm}])$. The calibration is valid for
stars with $4500<T_{\mathrm{eff}}<6500$\,K, $K_P<14$ and flicker estimates
corresponding to stars with $3.25<\log g_{\star}<4.43$. Our 
relation has a model error in the stellar density of 31.7\% and so has $\sim8$ 
times lower precision than that from asteroseismology but is applicable to a 
sample $\sim40$ times greater. Flicker therefore provides an empirical method to 
enable asterodensity profiling on hundreds of planetary candidates from present 
and future missions.

\end{abstract}

% #####################################################################
%% keywords

%% EOF keywords
%% EOF titlepage

% #####################################################################
%% Introduction
\section{INTRODUCTION}
\label{sec:intro}

In recent years, there has been an increased interest in exploiting time series 
brightness variations of stars to infer fundamental stellar properties. This 
upsurge has been largely motivated by the transiting exoplanet survey missions,
such as CoRoT \citep{baglin:2006} and \emph{Kepler} \citep{borucki:2009}, which 
have provided an avalanche of high signal-to-noise and high cadence 
photometry. In addition, such missions require accurate stellar characterization 
to infer the correct parameters for the associated transiting planet candidates.

The strongest constraints on a star's fundamental parameters using time series 
photometry come from asteroseismology \citep{chaplin:2013}, the study of stellar 
oscillations. Giants, sub-giants and bright dwarfs provide data on solar-like 
oscillations, and in these cases fundamental parameters may be determined to the 
percent level. Another technique, known as gyrochronology, exploits a star's 
gradual angular momentum loss to date stars to $\sim15$\% accuracy using 
empirical calibrations from open clusters \citep{skumanich:1972,barnes:2007,
epstein:2014}. Here, the primary input is a stellar rotation period, which is 
revealed via rotational modulations in a photometric time series 
\citep{walkowicz:2013}. Recently, a new characterization technique has been 
proposed by \citet{bastien:2013} (B13), which uses the variability over an 
8-hour timescale, ``flicker'', as a proxy for the surface gravity 
of a star, $g_{\star}$. Flicker is able to reproduce $\log g_{\star}$ to within 
$\sim0.10$\,dex for FGK dwarfs and giants down to apparent magnitudes of 14 and 
is thought to be physically caused by stellar granulation on the star's surface 
\citep{mathur:2011,cranmer:2014}. Whilst asteroseismology undoubtedly provides 
tighter constraints on a star's basic parameters, the ability of flicker to 
infer $\log g_{\star}$ for a much larger number of stars in a magnitude-limited 
survey, such as \emph{Kepler}, makes it highly appealing. 

In conjunction with these recent developments in stellar characterization, 
several authors have recently explored how an accurate determination of the mean 
stellar density, $\rho_{\star}$, plus a high quality transit light curve may be
used to infer various properties of an exoplanet \citep{MAP:2012,
dawson:2012,AP:2014}. Asterodensity profiling (AP) compares the stellar density 
derived from the transit light curve shape \citep{seager:2003}, 
$\rho_{\star,\obs}$, to some independent measure, 
$\rho_{\star,\tru}$. Relative differences can be caused by numerous 
phenomena, including orbital eccentricity and blend scenarios \citep{AP:2014}. 
Whilst asteroseismology directly yields the mean stellar density for those 
targets with detected oscillations \citep{ulrich:1986}, flicker is currently 
only calibrated to surface gravity.

In this letter, we show that flicker is also able to determine the bulk density 
of a star to within $\sim30$\% across a wide range of spectral types
and apparent magnitudes. This new empirical relation opens the door to
conducting AP on many hundreds of transiting planet candidates detected by
both \emph{Kepler} and future missions. We describe our methodology for
deriving this relation in \S\ref{sec:method}, followed by an exploration of the
results in \S\ref{sec:results}. We close in \S\ref{sec:discussion} by discussing
the potential of this relation for AP with both previous and future missions.

\section{METHODS}
\label{sec:method}

\subsection{Seismic Data}
\label{sub:seismicdata}

In order to investigate whether a relation exists between flicker and mean
stellar density, we first require an accurate catalog of stellar densities. 
Following B13, we identify the sample of \emph{Kepler} targets for which 
asteroseismology oscillation modes have been detected as our ``gold standard'' 
catalog. These asteroseismology detections provide two immediate basic 
parameters: the average large frequency separation between consecutive overtones 
of the same spherical angular degree, $\Delta\nu$, and the frequency of maximum
oscillations power, $\nu_{\mathrm{max}}$. This latter term has been shown
to be functionally dependent, to a good approximation, upon the surface gravity 
of the host star, $g_{\star}$, and the effective temperature, 
$T_{\mathrm{eff}}$, \citep{brown:1991,kjeldsen:1995,chaplin:2008,belkacem:2011} 
via:

\begin{align}
\big( \nu_{\mathrm{max}} / \nu_{\mathrm{max},\odot} \big) \simeq \big(g_{\star}/g_{\odot}\big) \big(T_{\mathrm{eff}}/T_{\mathrm{eff},\odot} \big)^{-1/2},
\end{align}

where $\nu_{\mathrm{max},\odot}$ is Sun's frequency of maximum oscillations 
power, $T_{\mathrm{eff},\odot}$ is the Sun's effective temperature and 
$g_{\odot}$ is the Sun's surface gravity. The derived surface gravity is 
therefore moderately dependent upon some independent measure of the effective 
temperature. The other basic seismic observable, $\Delta\nu$, scales with the 
star's mean stellar density \citep{ulrich:1986} as:

\begin{align}
\big( \Delta\nu / \Delta\nu_{\odot} \big) \simeq \sqrt{ \rho_{\star} / \rho_{\odot} },
\end{align}

where $\rho_{\odot}$ is the Sun's mean density. In practice, this scaling is 
only approximate, and comparisons to individual model frequencies have shown 
temperature-dependent offsets of up to 2\% in $\Delta\nu$ \citep{white:2011}. 
Nevertheless, $\rho_{\star}$ has a weaker functional dependence on the input 
effective temperature than $g_{\star}$ and detailed seismic modeling allows for 
a refined measurement of $\rho_{\star}$ at the percent level. For this reason, 
we consider that the seismic densities are the most accurate and precise 
parameter revealed by asteroseismology and so provide an ideal catalog for later 
calibration to flicker estimates.

In this work, we use the catalogs of \citet{huber:2013} and 
\citet{chaplin:2014}, which include 588 distinct \emph{Kepler} target stars.
In the case of the catalog from \citet{chaplin:2014}, three different estimates
of $\rho_{\star}$ are available, which vary very slightly due to different
assumed effective temperatures and metallicities (Tables 4, 5 \& 6). Of these, 
we preferentially employ Table~6 values wherever available (using high resolution 
spectroscopic stellar inputs from \citealt{bruntt:2012}), then Table~5 values
(IRFM $T_{\mathrm{eff}}$ and field-average metallicity values) and finally 
Table~4 in all other cases (using SDSS-calibrated $T_{\mathrm{eff}}$ and 
field-average metallicity values). In the few cases where both 
\citet{huber:2013} and \citet{chaplin:2014} have independent measures, we use
\citet{huber:2013} due to their use of dedicated spectroscopic inputs. In all
cases, $\rho_{\star}$ uncertainties are also available, to provide the relevant 
weightings in the later regressions (see \S\ref{sub:deming}).

\subsection{Flicker Data}
\label{sub:flickerdata}

Following \citet{basri:2011}, and further described in B13, we measure the 
high-frequency stellar noise ($F_8$) in the standard pipeline processed PDC-MAP
(data release 21) \emph{Kepler} long-cadence (30\,min) light curves by 
calculating the root-mean-square (RMS) of the difference between the light curve 
and a box-car smoothed version of itself. \citet{basri:2011} originally defined 
the high-frequency noise using a 4-point smoothing, however B13 found that a 
16-point (8-hr) smoothing yielded the cleanest correlation between the resultant 
$F_8$ and asteroseismically measured $\log g_{\star}$. Large excursions in the 
light curve, caused, for example, by stellar flares, can artificially inflate 
the measured $F_8$, resulting in erroneously low $F_8$-based $\log g_{\star}$. 
As such, we clip all $2.5$\,$\sigma$ or greater outliers in the light curve 
prior to taking the RMS. For stars with known transiting exoplanets, such as 
those in \citet{huber:2013}, we additionally excise in-transit data points using 
the planetary orbital parameters publicly available through the NASA Exoplanet 
Archive \citep{akeson:2013}. We then correct this derived $F_8$ for shot noise 
contributions as described in B13. We compute $F_8$ for all available quarters 
of \emph{Kepler} data and adopt our final estimate and associated uncertainty as 
the mean and standard deviation of the results, respectively. 

\subsection{Deming Regression}
\label{sub:deming}

We here describe how we compute the ``best-fitting'' model describing the
dependent variable, $\rho_{\star}$, with respect to the independent variable,
$F_8$. We begin by noting that the correlation between these terms is close to 
linear using log-log scaling (see Fig.~\ref{fig:relation}), similar to the case 
when B13 compared $\log g_{\star}$ to $F_8$. Switching to log-log scaling 
requires an adjustment of the uncertainties in both variables. Uncertainties are 
known for both the $\rho_{\star}$ (see \S\ref{sub:seismicdata}) and $F_8$ 
(see \S\ref{sub:flickerdata}) measurements, which may be converted into 
$\log_{10}$ space using the general rule:

\begin{align}
\sigma_{\log_{10} z} &= (\sigma_z/z)/\log_{e} 10,
\end{align}

where $z$ is each variable and $\sigma_z$ is the associated uncertainty. Next,
we consider that any model employed will itself be somewhat erroneous; i.e. 
there are not only uncertainties on the observables, but also in the model 
itself. For example, this could be because other terms not considered here
also impact the observed flicker. We adopt a simple method to treat the 
model uncertainty by introducing a quadrature error term in the dependent
variable, $\log_{10}\rho_{\star}$, given by $\sigma_{\mathrm{model}}$. This is 
similar to the way in which stellar ``jitter'' is often treated in radial 
velocity regressions \citep{wright:2005} and implicitly assumes that the model 
error does not vary with respect to the independent variable.

Although the fractional uncertainties on the $\log_{10} F_8$ measurements are 
typically much greater than those of $\log_{10} \rho_{\star}$, we seek a 
regression technique which accounts for the appropriate weighting in both 
observables. Accordingly, our regression is performed in the least-squares
framework, but specifically using the generalized \citet{deming:1943} method,
which accounts for the uncertainties in both variables. We adopt a simple linear 
slope model given by Equation~\ref{eqn:model}, motivated by the visual 
correlation (see Fig.~\ref{fig:relation}) and the simplicity of this model 
makes it attractive for wider use in the community.

\begin{align}
\log_{10}\big(\rho_{\star}\,[\mathrm{kg}\,\mathrm{m}^{-3}]\big) = \alpha + \beta \log_{10}\big(F_8\,[\mathrm{ppm}]\big).
\label{eqn:model}
\end{align}

To perform a Deming regression, we must determine the co-ordinates
of the point along the model curve, $y(x)$ (where $y$ and $x$ denote the 
dependent and independent variables respectively), which has the closest 
Euclidean distance to each trial point, $\{x_i,y_i\}^T$. This is achieved by 
minimizing the metric $(x_{c,i}-x_i)^2 + (y(x_{c,i})-y_i)^2$ with respect to 
$x_{c,i}$, where $\{x_{c,i},y_{c,i}\}^T$ is the point along the curve $y(x)$ 
closest to the trial point $\{x_i,y_i\}^T$, which for our linear slope model 
gives:

\begin{align}
%x_{c,i} &= \frac{\alpha \beta + x_i + \beta y_i}{1+\beta^2},\\
%y_{c,i} &= \frac{\alpha + \beta (x_i + \beta y_i)}{1+\beta^2}.
x_{c,i} &= (\alpha \beta + x_i + \beta y_i)(1+\beta^2)^{-1},\\
y_{c,i} &= (\alpha + \beta (x_i + \beta y_i))(1+\beta^2)^{-1}.
\end{align}

The least-squares merit function, which we numerically minimize with respect to 
$\alpha$ and $\beta$, is simply the weighted Euclidean distance between the 
observations and the model:

\begin{align}
\sum_{i=1}^N \frac{ (x_i - x_{c,i})^2 + (y_i - y_{c,i})^2 }{ \sigma_{x,i}^2 + \sigma_{y,i}^2},
\label{eqn:meritfn}
\end{align}

where in our specific case we have $x_i\to
\log_{10}\big(F_8\,[\mathrm{ppm}]\big)$, $y_i\to
\log_{10}\big(\rho_{\star}\,[\mathrm{kg}\,\mathrm{m}^{-3}]\big)$,
$\sigma_{x,i}^2\to\sigma_{\log_{10}(F_8\,[\mathrm{ppm}])}^2$ and
$\sigma_{y,i}^2\to
\sigma_{(\rho_{\star}\,[\mathrm{kg}\,\mathrm{m}^{-3}])}^2 + 
\sigma_{\mathrm{model}}^2$.

If we assume that the residuals of the dependent variable to the best-fitting 
model are approximately normally distributed, then one would expect, for a
well-chosen $\sigma_{\mathrm{model}}$, that the residuals divided by the
respective uncertainties would be well described by the standard normal
distribution. This can be checked by a subsequent least-squares regression of a 
normal distribution of zero mean but freely fitted variance to the distribution 
of the residuals. In practice, this is performed on the cumulative distribution 
of the residuals, rather than a probability density histogram, to avoid the 
choice of the bin-size affecting the results \citep{beta:2013}. In 
general, an arbitrary guess of $\sigma_{\mathrm{model}}$ will lead to a 
non-unity variance for this regressed normal distribution, but we iterate 
$\sigma_{\mathrm{model}}$ until this condition is satisfied in order to solve 
for the best $\sigma_{\mathrm{model}}$.

\section{Results}
\label{sec:results}

\subsection{Best-fitting Model}
\label{sub:fullresults}

\begin{figure*}
\begin{center}
\includegraphics[width=16.8 cm]{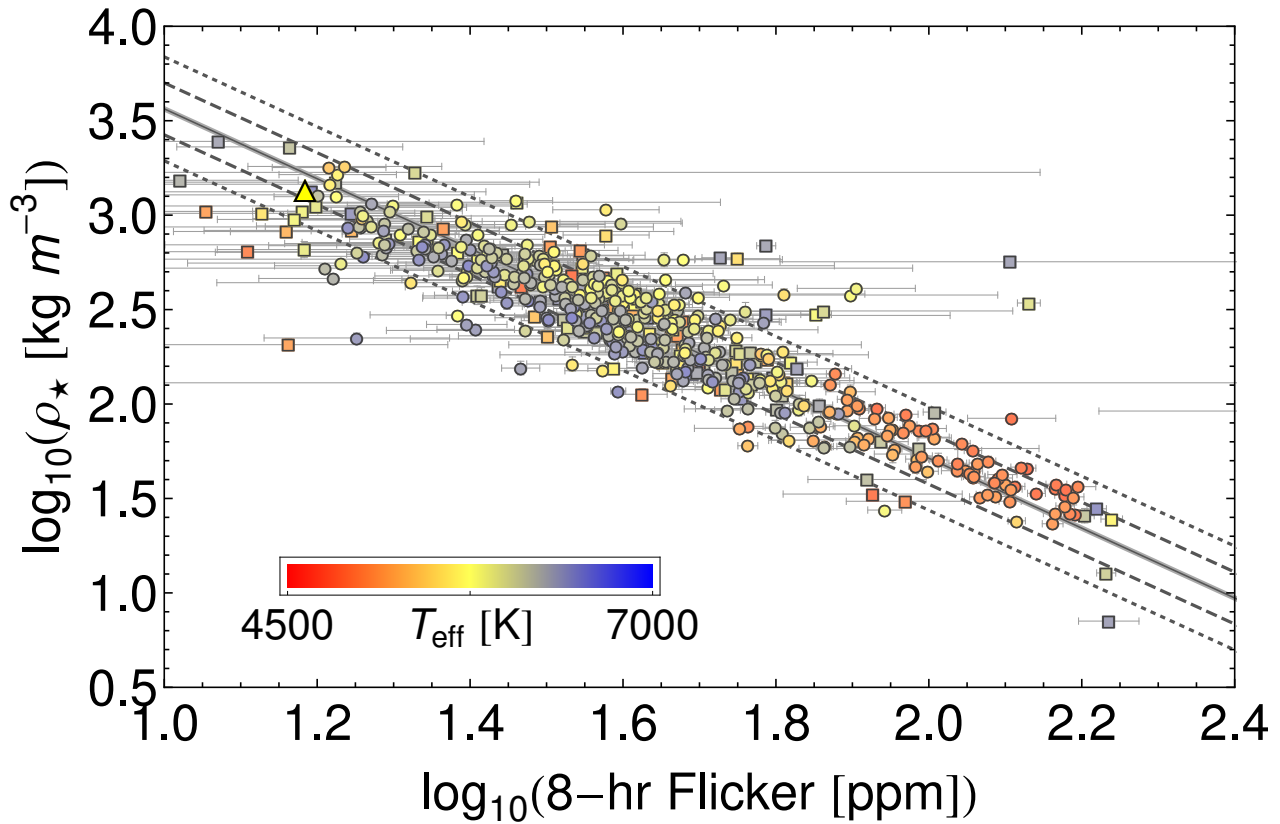}
\caption{\emph{Empirical relationship between stellar density, $\rho_{\star}$,
and the 8-hour flicker, $F_8$. Squares are unreliable points not used in 
regressing the best-fitting linear relation (solid line). Dashed 
and dotted lines show the 1\,$\sigma$ and 2\,$\sigma$ confidence regions.
Points are color coded by the effective temperature and the Sun is marked with
a triangle.
}} 
\label{fig:relation}
\end{center}
\end{figure*}

Following the method described in \S\ref{sub:deming}, the best-fitting (mean 
maximum likelihood) model is shown in Fig.~\ref{fig:relation}. Our simple 
linear model provides an excellent description of the apparent correlation and 
the associated parameters are available in Table~\ref{tab:params}. To compute 
this model, we enforced several criteria to exclude some of the less reliable 
data.

\begin{itemize}
\item[{\tiny$\blacksquare$}] $1.2<\log_{10}\big(F_8\,[\mathrm{ppm}]\big)\leq2.2$
since we have relatively few points outside of this range.
\item[{\tiny$\blacksquare$}] Range$\leq1000$\,ppm (defined in B13), since 
B13 find these points to be more frequently outliers.
\item[{\tiny$\blacksquare$}] $4500\leq T_{\mathrm{eff}}<6500$\,K to avoid
the very cool or hot stars in our sample.
\item[{\tiny$\blacksquare$}] $K_P<14$ since correcting $F_8$ for 
\emph{Kepler} apparent magnitude is only calibrated up to this level (B13).
\end{itemize}

Note that the first criterion is equivalent to 
$3.25\lesssim\log g_{\star}\lesssim4.43$ using the B13 relation, thus
eliminating M \& K dwarfs with insufficient photometric variability.
These filters reduce the number of data points from 588 to 439. The
439 ``good'' points are denoted by circles in Fig.~\ref{fig:relation} and the
ignored ``bad'' points are denoted by squares. Using only the good
points, we find the model error is 0.14\,dex (32\% in $\rho_{\star}$). 
Including all the data increases this to 0.15\,dex (35\% in $\rho_{\star}$)
and the fitted parameters are provided in Table~\ref{tab:params}.

\begin{table}
\caption{\emph{Best-fitting parameters between our linear model and the
$\rho_{\star}$-$F_8$ data for two sample choices.}} % title of Table
\centering % used for centering table
\begin{tabular}{c c c c c} % centered columns (4 columns)
\hline\hline %inserts double horizontal lines
& $\alpha$ & $\beta$ & $\sigma_{\mathrm{model}}$ & Frac. Err. in $\rho_{\star}$ \\ [0.5ex] % inserts table
%heading
\hline % inserts single horizontal line
``Good'' sample & $5.413$ & $-1.850$ & $0.138$ & $31.7$\% \\
``Full'' sample & $5.471$ & $-1.884$ & $0.151$ & $34.7$\% \\ [1ex]
\hline\hline %inserts single line
\end{tabular}
\label{tab:params} % is used to refer this table in the text
\end{table}

\begin{figure}
\begin{center}
\includegraphics[width=8.4 cm]{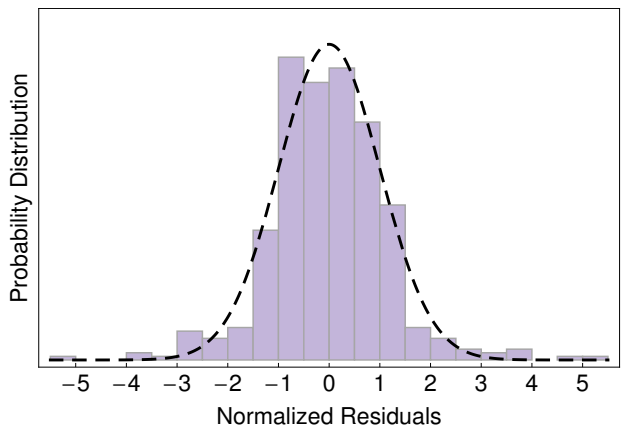}
\caption{\emph{Distribution of the residuals of 
$\log_{10}(\rho_{\star}\,[\mathrm{kg}\,\mathrm{m}^{-3}])$ to best-fitting
linear model for the ``good'' sample, where we have normalized each residual by
the associated uncertainty. The black dashed line is a standard normal 
distribution.
}} 
\label{fig:histo}
\end{center}
\end{figure}

Fig.~\ref{fig:histo} reveals that the residuals of the ``good'' sample appear
approximately normally distributed. This means that if one knows the flicker of
a star, one may construct an informative prior on $\log\rho_{\star}$ using a
normal distribution and the parameters provided in Table~\ref{tab:params}.
For the Sun's observed flicker (B13), our relation yields
$\rho_{\star}=(1.18\pm0.37)$\,$\rho_{\odot}$, compatible with the truth.

\subsection{Comparison to the Surface Gravity Relation}
\label{sub:loggresults}

We repeated our regression on the ``good'' sample replacing 
$\log_{10}(\rho_{\star}\,[\mathrm{kg}\,\mathrm{m}^{-3}])$ with
$\log_{10}(g_{\star}\,[\mathrm{cm}\,\mathrm{s}^{-2}])$. Since the dependent
variables have been changed and use different units, a fair comparison of their
relative performance cannot be conducted by inspection of the residuals in
the dependent variables. Instead, we invert the best-fitting model to predict
$F_8$ as a function of the dependent variable and then compute the residuals
in the $F_8$ measurements between the two models.

We find that the $\rho_{\star}$ model has a standard deviation in 
$\log_{10}(F_8\,[\mathrm{ppm}])$ of $0.0866$\,dex, versus $0.0824$\,dex for the
$\log g_{\star}$ model. We also note that the Pearson's correlation coefficient 
is $-0.927$ for the $\rho_{\star}$ model but $-0.934$ for the $\log g_{\star}$ 
model. We therefore conclude that surface gravity is a slightly better dependent 
variable to correlate against $F_8$, although clearly $\rho_{\star}$ does an 
excellent job too (see Fig.~\ref{fig:relation}). This weakly supports the 
hypothesis that flicker is most directly tracing surface features such as 
granulation \citep{cranmer:2014}, rather than internal processes. Further 
support of this is found when we compare $M_{\star}/R_{\star}$ to $F_8$, causing 
the correlation coefficient to drop to $-0.928$. The $\rho_{\star}$ relation is 
likely a good match simply due to the expected evolutionary correlation between 
$M_{\star}/R_{\star}^2$ and $M_{\star}/R_{\star}^3$. Nevertheless, the 
$\rho_{\star}$ relation is more useful when analyzing exoplanet transits, 
since this term directly affects the light curve shape \citep{seager:2003}.

\section{Discussion}
\label{sec:discussion}

\subsection{Flicker as an Input for Asterodensity Profiling}
\label{sub:AP}

Asterodensity profiling (AP) has recently emerged as a valuable tool for
characterizing exoplanets using time series photometry
\citep{MAP:2012,sliski:2014,AP:2014}. AP exploits the fact that for a planet
on a Keplerian circular orbit transiting an unblended star with a symmetric 
intensity profile, the shape of the light curve reveals $\rho_{\star,\obs}$ 
\citep{seager:2003}. If any of the idealized assumptions are
invalid, then $\rho_{\star,\obs}$ will differ from the true value, 
$\rho_{\star,\tru}$, and the direction and magnitude of the discrepancy reveals
information about the transiting system \citep{AP:2014}. Using independent 
$\rho_{\star}$ estimates from asteroseismology, \citet{sliski:2014}
provide an example of the utility of AP by showing that the false positive
rate of transiting planet candidates associated with giant stars is much higher
than that of dwarf stars. 

In this work, we have shown that flicker may also be used as an input for the 
independent measure of $\rho_{\star,\tru}$ required for AP. Despite 
uncertainties in $\rho_{\star}$ increasing to $\sim30$\% from a flicker-based 
determination versus $\sim4$\% using asteroseismology, flicker can be used on 
many more targets in a magnitude-limited photometric survey like \emph{Kepler}, 
since it works reliably down to $K_P=14$ (see later discussion in 
\S\ref{sub:implications}). We do not claim that the derived relation is the
optimal choice of regressors or parametric form, merely that it provides
a simple, empirical recipe for estimating $\rho_{\star}$. For example, including
effective temperature may improve the relation, since cooler stars seem to be 
found at higher flicker values (see Fig.~\ref{fig:relation}). However, including
such terms would make our relation no longer purely photometric, which we would
argue is the principal benefit of the flicker technique.

\begin{figure}
\begin{center}
\includegraphics[width=8.4 cm]{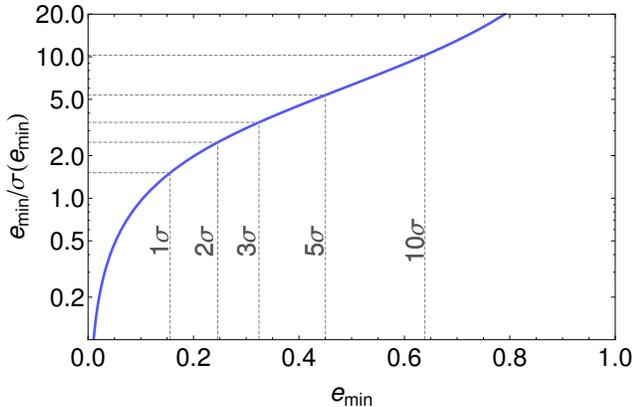}
\caption{\emph{Sensitivity of flicker to detecting eccentric exoplanets via
the photo-eccentric effect. We here assume a fractional error in 
$\rho_{\star,\tru}$ from a flicker-based measurement of 31.7\%. Several key
confidence levels are marked with dashed grid lines for reference.
}} 
\label{fig:eminplot}
\end{center}
\end{figure}

Whilst we direct those interested to \citet{AP:2014} for details on the theory
and range of effects which can cause AP discrepancies, we here provide an
example calculation of the sensitivity of AP using flicker to detect eccentric
exoplanets via the so-called ``photo-eccentric'' effect \citep{dawson:2012}.
For a planet on an eccentric orbit, the derived light curve stellar density
will differ from the true value by \citep{investigations:2010}:

\begin{align}
\Big( \frac{\rho_{\star,\obs}}{\rho_{\star,\tru}} \Big) &= \frac{(1+e\sin\omega)^{3}}{(1-e^2)^{3/2}},
\label{eqn:psi}
\end{align}

where $e$ is the orbital eccentricity and $\omega$ is argument of periastron.
\citet{dawson:2012} show how to first order, constraints on $e$ scale with
$\rho_{\star,\tru}^{1/3}$ and thus even a weak prior on the density can
lead to useful constraints on $e$. With one observable and two 
unknowns, a unique solution to Equation~\ref{eqn:psi} is not possible. However, 
one can derive the minimum eccentricity, $e_{\mathrm{min}}$, of the planet and 
the associated uncertainty, $\sigma_{e_{\mathrm{min}}}$, using Equations~39\&40 
of \citet{AP:2014} respectively. Let us assume that the uncertainty in 
$(\rho_{\star,\obs}/\rho_{\star,\tru})$ is dominated by the denominator's
error, which in turn was found using our flicker relation and equals 
31.7\%. We may now plot the term 
$(e_{\mathrm{min}}/\sigma_{e_{\mathrm{min}}})$ as a function of 
$e_{\mathrm{min}}$ in Fig.~\ref{fig:eminplot}, to illustrate the 
ability of flicker to detect eccentric planets. Using the classic 
\citet{lucy:1971} test, we mark several key confidence levels with grid lines,
demonstrating that flicker can detect eccentricities of 
$e_{\mathrm{min}}=0.25$ to $\geq2$\,$\sigma$ confidence and 
$e_{\mathrm{min}}=0.32$ to $\geq3$\,$\sigma$. With the power of large number
statistics, we anticipate that flicker will be particularly powerful for 
inferring the ensemble distribution of orbital eccentricities.

\subsection{Implications}
\label{sub:implications}

Between the two catalogs of \citet{huber:2013} and \citet{chaplin:2014}, there 
are 588 unique targets with asteroseismology detections yielding $\rho_{\star}$ 
measurements with a median uncertainty of 4.1\%. In contrast, there are 
$28,577$ \emph{Kepler} targets with $K_P<14$, $4500<T_{\mathrm{eff}}<
6500$\,K and $3.25<\log g<4.43$ (NASA Exoplanet Archive). Making the simple 
assumption that the same fraction of these targets will satisfy the range 
criterion defined earlier in \S\ref{sub:fullresults}, then we expect 
$\sim25,000$ targets to be amenable for a flicker-based estimate of 
$\rho_{\star}$ with a model accuracy of 31.7\%. This translates to an increase 
in the number of AP targets by a factor of $\sim40$ at the expense of an 
increase in the measurement uncertainty by a factor of $\sim8$. By any accounts, 
this is an acceptable compromise and opens the door to conducting AP on hundreds 
of \emph{Kepler} planetary candidates (we estimate $\sim630$).

An alternative method to determine $\rho_{\star,\tru}$ for targets without
detectable oscillations would be via spectroscopy (e.g. \citealt{dawson:2012}). 
Here, one observes a spectrum of the target, compares it to a catalog of library 
spectra with various $T_{\mathrm{eff}}$, $[\mathrm{M}/\mathrm{H}]$ and 
$\log g_{\star}$ and then finally one finds the best matching stellar evolution 
isochrones to these basic parameters. This procedure has several drawbacks 
compared to a flicker-based determination though. Firstly, this method requires 
that one obtain high SNR spectra, whereas F8 can be measured from the data 
obtained directly from a photometric mission like \emph{Kepler}. Secondly, the 
final determination of $\rho_{\star}$ is strongly model dependent using both 
stellar evolution models and spectra template matching laden with challenging 
degeneracies. Finally, it is worth noting that the formal uncertainty on a 
spectroscopic determination of $\rho_{\star,\tru}$ is typically no better than 
the flicker-based empirical relation for Sun-like stars. For example, 
\citet{dawson:2012} report 
$\rho_{\star,\tru}=1.02_{-0.29}^{+0.45}$\,$\rho_{\odot}$ for the $K_P=13.6$ 
Sun-like target KOI-686 and our flicker technique yields 
$(0.97\pm0.44)$\,$\rho_{\odot}$. In general then, we argue that for 
determinations of $\rho_{\star}$, the empirical and largely model 
independent flicker technique is preferable to spectroscopy, provided the target 
satisifies our sample criteria.

For the future TESS mission \citep{ricker:2010}, the smaller lens aperture of 
12\,cm will lead to higher photon noise than \emph{Kepler}, for the same target. 
We therefore expect the 14$^{\mathrm{th}}$ magnitude cut-off of our flicker 
calibration to drop to $\sim11.5$. The same effect will lead to only very 
bright stars having asteroseismology detections though, with preliminary 
estimates suggesting $\sim5\times10^3$ asteroseismology targets out of 
$\sim5\times10^5$ target stars. In contrast, we expect that $\sim10^5$ TESS 
targets will be amenable to a flicker-based determination of their stellar 
densities (with the exact number depending upon the as yet unknown target list). 
Similarly, we expect flicker to have majorly benefit the upcoming PLATO 
2.0 mission \citep{rauer:2013} for both moderately bright targets near the edge
of the field and faint targets in the center. Since AP is not only a method for 
characterizing exoplanets but also for vetting them \citep{sliski:2014}, then 
we expect flicker to be an invaluable tool in the TESS and PLATO era.

% #####################################################################
%% Acknowledgements
\acknowledgements
\section*{Acknowledgements}

% Personal funding
DMK is supported by the NASA Sagan Fellowships. FAB is supported by the 
NASA Harriet Jenkins Fellowship and a Vanderbilt Provost Graduate Fellowship.
WJC acknowledges financial support from the UK Science and Technology Facilities 
Council. DH acknowledges support by an appointment to the NASA Postdoctoral 
Program at Ames Research Center, administered by Oak Ridge Associated 
Universities through a contract with NASA, and support by the Kepler 
Participating Scientist Program.
%
%% EOF Acknowledgements
\clearpage
% #####################################################################
%% Bibliography

\end{document}